\documentstyle[aps,epsf]{revtex}
\begin{document}

\preprint{HEP/123-qed}

\title{Non-classical interference between independent sources}

\author{J.G.Rarity, P.R.Tapster and R.Loudon}

\address{DRA Malvern, \\St.Andrews Rd, Malvern, \\Worcs., UK,
WR14 3PS}

\date{\today}
\maketitle

\begin{abstract}
When a one-photon state is mixed with a (separate) weak coherent state at a
beamsplitter the probability for seeing one photon in each beamsplitter
output approaches zero due to destructive interference. We demonstrate this
non-classical interference effect using pulse-gated single photons and weak
mode-locked laser pulses.
\end{abstract}


Entanglement and non-classical interference effects of particles with no
previous history in common has been the subject of several recent theoretical
studies\cite{Yurke92,ZeEkWe94,Rarity95}.  As well as providing further
evidence in support of the conventional view of quantum mechanics the
realisation of such experiments will prove the feasibility of entangling the
large numbers of quanta essential for quantum computation \cite{QComp}.  Here
we perform the simplest experiment capable of showing non-classical
interference from independent sources by mixing a one-photon state and a weak
coherent state at a beamsplitter.  This is essentially a reduced version of a
heterodyne experiment designed to show non-local interference first suggested
by Tan et al \cite{Tan89}.  When the weak coherent state cannot be
distinguished from the one-photon state by measurement after the beamsplitter
we show that the probability of a coincident detection across the beamsplitter
output ports goes to zero.  In a comparable experiment using classical beams
from independent sources \cite{Pfleegor} the random phase fluctuations lead to
a time varying fringe pattern.  In the limit where the time uncertainty
(jitter) in detection is much less than the rate of phase drift a similar
coincidence suppression will be seen but with reduced visibility due to
averaging over the random relative phases.  In the experiment we derive a good
approximation to a one-photon Fock state by using a pulsed parametric
downconversion source emitting time correlated photon-pairs.  Detection of one
photon of the pair is used to gate detection of its partner \cite{RaTaJa87} to
within a time window given by the short time duration of the pump pulse (a few
hundred femto-seconds).  The pump for parametric downconversion is obtained by
frequency doubling a femtosecond pulsed laser.  A strongly attenuated
component of the original laser beam is used as a weak coherent state.  This
source has no phase correlation with individual beams of the downconverted
light and can be thought of as an independent source.  The weak coherent state
and one photon state can be made indistinguishable by making their time
uncertainty longer than their initial pulse lengths by filtering in a
narrowband filter before detection.  We can investigate the interference
between independent classical pulsed sources in the same experiment by
accepting all downconversion events (ignoring the gating).

In figure 1 we illustrate a lossless beamsplitter with amplitude transmission
and reflection coefficients $t=|t|e^{i\phi_t}$ and $r=|r|e^{i\phi_r}$.  Energy
conservation requires $r^*t+rt^*=0$ and $|r|^2+|t|^2=1$ \cite{Loudon87} thus
$\phi_t-\phi_r=\pi/2$.  The single modes at the input ports
$|\hspace{1mm}\rangle_a,|\hspace{1mm}\rangle_b$ of the beamsplitter are
populated by creation operators $a^\dagger$ and $b^\dagger$ respectively.
Then the beamsplitter transforms $a^\dagger$ and $b^\dagger$ into the creation
operators $c^\dagger$ and $d^\dagger$ at the output ports of the beamsplitter
\cite{Loudon87,Mandel87,Campos89} through
$a^\dagger\rightarrow t c^\dagger + r d^\dagger$ and $ b^\dagger \rightarrow
r c^\dagger + t d^\dagger$.  Taking a vacuum state at the $b$-port
$|0\rangle_b$ and a number state $|n\rangle_a$ at the $a$-port we obtain the
well known result of a binomial distribution for the probability of finding
$j$ photons in the $c$-mode and $n-j$ photons in the $d$-mode
$P_{cd}(j,n-j)=t^{2j} r^{2(n-j)}\;n!/(j!(n-j)!)$ This result would be obtained
if we took a simple `coin tossing' picture where each input `photon' takes a
random output direction independent of all other photons.

This picture becomes invalid when we populate each input with a single photon
state
\begin{equation}
|1\rangle_a |1\rangle_b \rightarrow (t^2+r^2)|1\rangle_c |1\rangle_d +
rt\sqrt{2}(|2\rangle_c|0\rangle_d +|0\rangle_c |2\rangle_d).
\end{equation}
When $|r|=|t|=1/\sqrt 2$ (the 50/50 beamsplitter) $t^2+r^2=0$ and the
probability of detecting one photon at each output port is zero
\cite{Loudon87} which is clearly non-classical.  Using the simple `coin
tossing' picture of beamsplitters this result appears counter-intuitive as
there is only one `coin' tossed for each pair.  This effect was first seen
when the coincident photon pairs created in parametric downconversion were
mixed at a beamsplitter\cite{Hong87,Us88}.

Generalising to arbitrary number states at both inputs to the beamsplitter we
write
\begin{equation}
 |n\rangle_a |m\rangle_b \rightarrow
 (t c^\dagger + r d^\dagger)^n  (r c^\dagger + t d^\dagger)^m
 |0\rangle_c|0\rangle_d / \sqrt{n!m!}
\end{equation}
\[
  =\sum_{j=0}^n\sum_{k=0}^m
   t^{m-k+j} r^{n-k+j} \sqrt{^nC_j\;^mC_k\;^{(j+k)}C_j\;
  ^{(n+m-j-k)}C_{(n-j)}}\: |(j+k)\rangle_c |(n+m-j-k)\rangle_d
\]

\begin{figure}
\epsfxsize=17cm
\epsffile{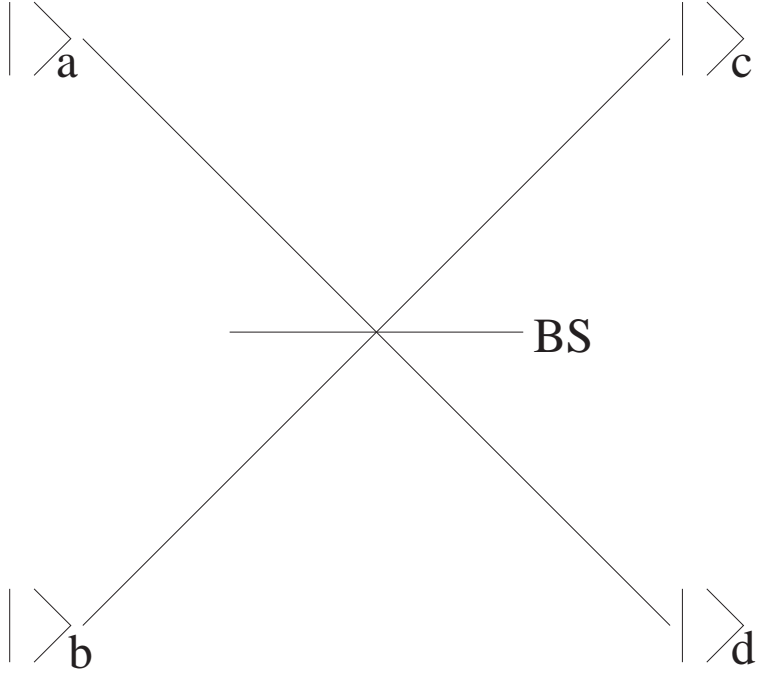}
\caption{ Overlapping modes at a beamsplitter (BS)}
\label{BS}
\end{figure}

Here we aim to demonstrate interference and thus entanglement between separate
sources rather than correlated photon pairs.  In the experiment we mix a one
photon state with an approximate coherent state $ |\alpha\rangle$, where the
mean number of photons in the coherent state is $ |\alpha|^2 $.  We can
express the coherent state as a linear combination of number states
$ |\alpha\rangle_a= \exp(-|\alpha|^2/2) \sum_{n=0}^\infty{\alpha^n /
\sqrt{n!}}|n\rangle_a $. Applying equation (2) the outputs can be expressed as
\begin{equation} \begin{array}{rl}
  |\alpha\rangle_a|1\rangle_b \rightarrow \exp(-|\alpha|^2/2)&
   \{ r |1\rangle_c |0\rangle_d + t |0\rangle_c |1\rangle_d + \\
  & \alpha [ (t^2+r^2)|1\rangle_c |1\rangle_d + 
  rt\sqrt{2}(|2\rangle_c |0\rangle_d +|0\rangle_c |2\rangle_d)] + \\
  & \alpha^2 [  rt^2\sqrt{3}|3\rangle_c |0\rangle_d +
    t(t^2+2r^2)|2\rangle_c |1\rangle_d + \\
  &  r(2t^2+r^2)|1\rangle_c |2\rangle_d +tr^2\sqrt{3}|0\rangle_c |3\rangle_d ]
    /\sqrt{2}+O(\alpha^3)\}
\end{array}\end{equation}
to second order in $\alpha$.

In the case where the beamsplitter is 50/50 equation 3 again predicts that the
probability of seeing exactly one photon in each port of the beamsplitter is
zero, independent of $\alpha$.  Realisable photon counting detectors will fire
once when they see one or more photons, thus this aspect of the non-classical
behavior will be masked when terms in $|n\rangle_c|m\rangle_d$; $n,m > 1$ are
present.  However these probabilities become negligible as $\alpha$ is reduced
to the point where we can ignore powers of $\alpha^2$ and above.

Realistic detectors are also inefficient.  We model detectors with efficiency
$\eta$ as beamsplitters of transmission coefficient $\eta$ mixing the signal
with a vacuum state followed by an unit efficiency detector\cite{YuenShap80}.
As stated above the detectors will fire only once when one or more photons are
present thus we have four possible combinations of detections or
non-detections with both detectors.  For instance if the state reaching the
detectors is $|n\rangle_c |m\rangle_d$ we can write these four detection
probabilities as
\begin{equation}\begin{array}{l}
  P_{nm}(0,0) = (1-\eta_1)^n(1-\eta_2)^m\\
  P_{nm}(0,1) = (1-\eta_1)^n(1-(1-\eta_2)^m)\\
  P_{nm}(1,0) = (1-(1-\eta_1)^n)(1-\eta_2)^m\\
  P_{nm}(1,1) = (1-(1-\eta_1)^n)(1-(1-\eta_2)^m)
\end{array}\end{equation}

The probabilities of obtaining $|n\rangle_c|m\rangle_d$ given by the square
moduli of the coefficients in equation (3) can be summed with the above
weightings to give the total coincidence probability $P_{tot}(1,1)$.  This
contrasts with the case where there is no correlation when the $a$- and
$b$-modes do not overlap at the beamsplitter outputs.  Here the probability of
seeing a coincidence is just $P_{c}(1)P_d(1)$ the product of the probabilities
of a single count in detectors $c$ and $d$.  This can be obtained by separate
application of binomial partition to the two inputs but also from the
equations 4 above ($P_c(1)=P_{tot}(1,0)+P_{tot}(1,1)$ the singles rate is
invariant).  We define a visibility $V=1-P_{tot}(1,1)/P_c(1)P_d(1)$ which
measures the fractional reduction of the coincidence rate from its
uncorrelated value.  This is illustrated as a function of coherent state
intensity $|\alpha^2|$ in figure 2.  We see the interference visibility rise
to near unity as the intensity drops below one photon per pulse.

\begin{figure}
\epsfxsize=17cm
\epsffile{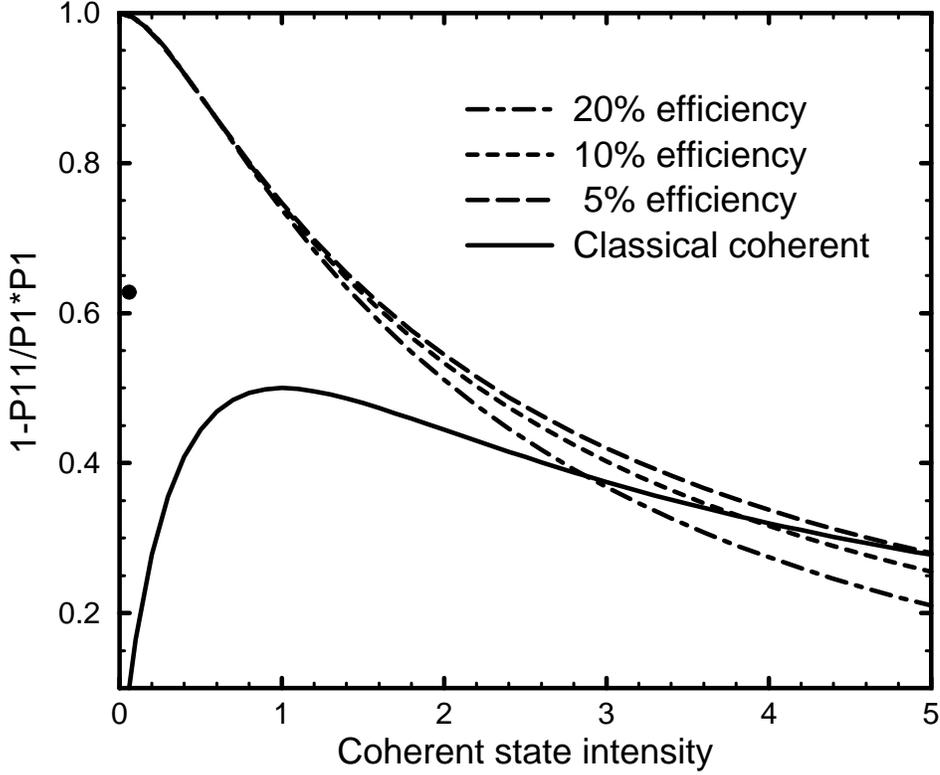}
\caption{ Interference visibilty between a coherent state and a
single photon state plotted as a function of the mean number of
photons in the coherent state $|\alpha|^2$ . Realistic detector
efficiencies are taken into account using equations 4. The solid
line shows the averaged interference visibility for two random
phase classical sources; the horizontal axis then represents
$R_{ab}$ and we hold $\langle I\rangle_b$ to be equivalent to one photon per
measurement. The experimental result is marked by a $\bullet$. }
\label{vis}
\end{figure}

We can compare this result with the classical case of two beams of intensity
$I_a$ and $I_b$ with randomly varying relative phase $\phi$ present at the
inputs of the beamsplitter.  The instantaneous $c$-mode output intensity is
given by
\begin{equation}
I_c=|t|^2I_a+|r|^2I_b-2|r||t|\sqrt{I_aI_b}\sin\phi
\end{equation}
and similarly in $d$.  When the intensities are low the coincidence rate
across the beamsplitter outputs is proportional to the product of output
intensities averaged over the random phase $\phi$
\begin{equation}
P(1,1)\propto \langle I_cI_d\rangle=|r|^2|t|^2(\langle I_a^2\rangle+
\langle I_b^2\rangle)+(|r|^4+|t|^4- 2|r|^2|t|^2)\langle I_a\rangle\langle I_b\rangle
\end{equation}
In the interference free case (separate modes) the subtractive term is
absent. The interference visibility using constant intensity sources and a
50/50 beamsplitter is thus
\begin{equation}
V=2R_{ab}/(R_{ab}+1)^2
\end{equation}
where $R_{ab}=\langle I_a\rangle/\langle I_b\rangle$. This is shown in figure
2 for comparison with the non-classical case assuming constant classical
intensity. Clearly the visibility reaches a maximum of 50\% when $I_a=I_b$.

In both classical and non-classical situations the visibility will be reduced
when the mode overlap is incomplete or when the phase fluctuates significantly
within the detection time. In previous experiments with separate classical
sources the phase drift was slow due to the narrowband nature of the laser
sources and standard detectors (10ns jitter) could be used. Here we limit the
detection time jitter by limiting the length of the light pulse itself and
use narrowband interference filters to ensure that the coherence time is
comparable or longer than this pulse length. We can perform multi-temporal
mode analysis of the non-classical interference effect when the coherent state
amplitude is small ($\alpha\ll 1$) by extending theory developed in
\cite{Rarity95,RaTa96}. We model the $c$- and $d$-mode filters by identical
Gaussians  with 1/$e^2$ intensity (half) width $\sigma$ and the doubled pump
beam as a time-bandwidth-product limited Gaussian pulse with $1/e^2$ intensity
width $\sigma_{2P}$ and assume the gating detector filter is much broader than
both. The probability per pulse of a triple coincidence between the gating
detector ($g$) and the $c$- and $d$-detectors is then given in the 50/50
beamsplitter case by
\begin{equation}
P(c,d,g)=\frac{1}{2}n_{dc}n_p \eta^3\left[1-V\exp\left(\frac{\Delta
X^2\sigma^2}{4c^2(1+\sigma^2/2\sigma_{2P}^2)}\right)\right]
\end{equation}
where $n_{dc}$ and $n_p$ ($\ll 1$) are the mean number per pulse of
downconverted pairs and weak coherent state photons before the beamsplitter
and $\eta^3$ is the product of the quantum efficiencies of the three
detectors. The coincidence rate as a
function of path length difference $\Delta X$ shows a Gaussian
dip with visibility $V=(1+\sigma^2/2\sigma_{2P}^2)^{-1/2}$ which
is unity when $\sigma\ll\sigma_{2P}$ and falls slowly as $\sigma$
increases.

\begin{figure}
\epsfxsize=17cm
\epsffile{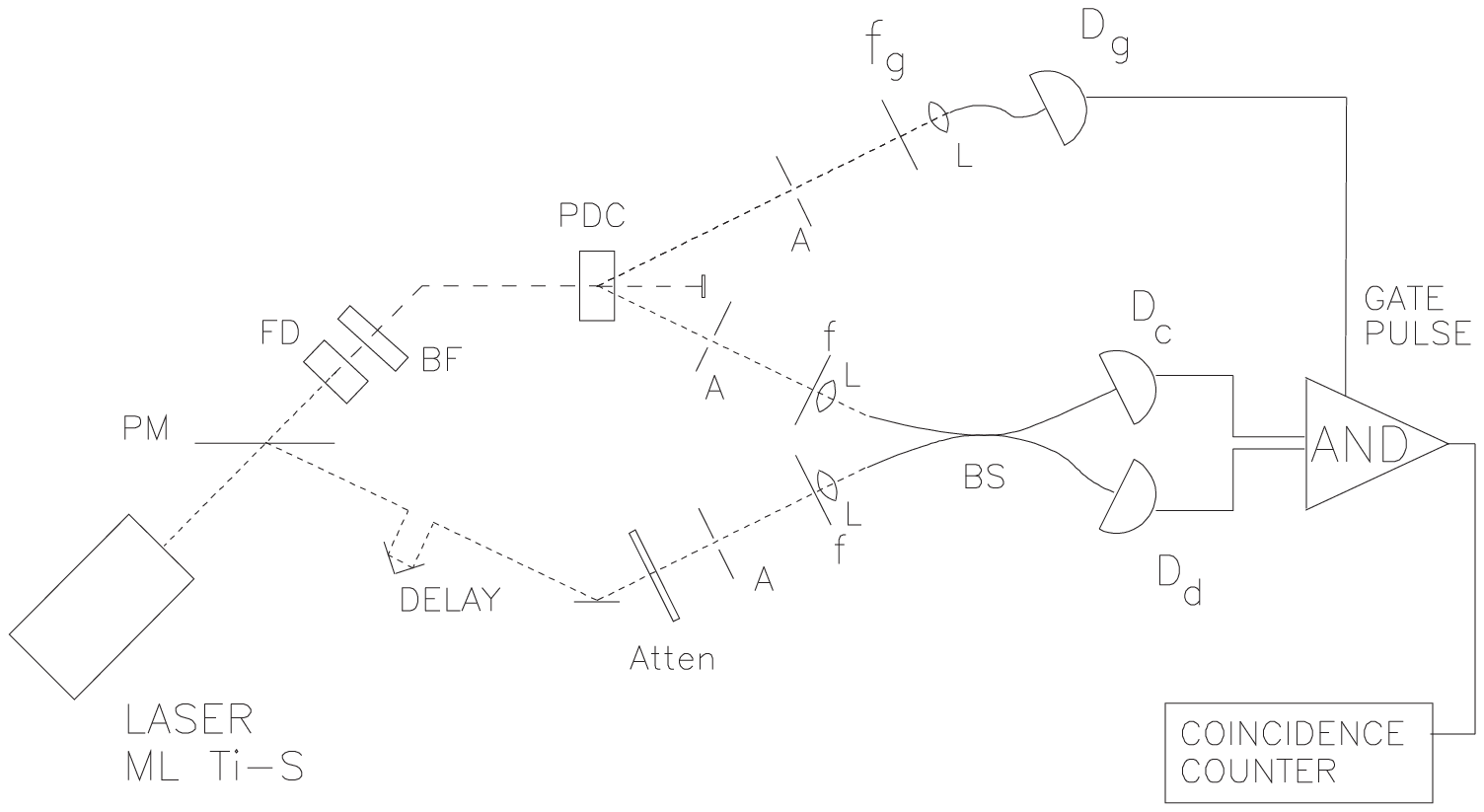}
\caption{The experimental apparatus. Key: (ML Ti-S) Mode locked
Ti-Sapphire laser, (PM) pick-off mirror, (FD) frequency doubler,
(BF) long wavelength blocking filter, (PDC) parametric
downconversion crystal, (A) apertures, (Atten) strong attenuator,
(f$_g$, f) interference filters with 815nm centre wavelength, (L)
lenses coupling light to optical fibres, (D$_{g,c,d}$) photon
counting avalanche photodiodes. Triple coincidences are measured
by a gated AND gate. Dotted lines indicate light of 815nm
wavelength, dashed lines indicate light of 407.5nm wavelength 
and thick curved solid lines represent optical fibres.}
\label{expt}
\end{figure}

The experimental apparatus is shown in figure 3.  A self mode locked
Ti-sapphire laser operating at 815nm centre wavelength ($\simeq$160fs pulse
length, 100MHz repetition rate) is doubled in a 2mm thick Beta Barium Borate
(BBO) crystal.  The resulting 407.5nm wavelength pulses are used to pump a 3mm
thick BBO crystal with crystal axis tilted to produce a cone of downconverted
light around 815nm with half angle $\simeq 7^\circ$.  Apertures placed at
opposite ends of a cone diameter ensure that we have a near ideal pulsed
pair-photon source centred on 815nm.  Optical fibres are used to couple this
light into photon counting detectors and to further restrict our source to a
single mode.  Long-wavelength-pass filters are used to ensure that the
detectors are not saturated by fluorescence and scatter from the violet pump
light.  After careful alignment a large proportion of detections are
coincident indicating effective detector efficiencies $\rangle$10\%
\cite{RaTaRi87}.  A photodetection in the gate arm of the apparatus is then
used to flag the presence of a single photon in the other arm to within the
pulse duration.  This single photon can then pass into a fibre coupler acting
as a 50/50 beamsplitter leading to two fibre coupled detectors (D$_c$, D$_d$).
Highly attenuated laser pulses are fed into the other input port of the
coupler.  By using a single mode fibre coupler we ensure that all the detected
downconversion and laser light is in a single mode and is indistinguishable
after the beamsplitter.  To minimise any dispersion effects the fibre pigtail
lengths to the coupler are cut equal to within a millimeter.  The
downconversion pulses and attenuated laser light can then be temporally
overlapped by adjusting a variable delay in the laser path length.  To further
ensure indistinguishability, 3nm bandwidth 815nm filters are placed before
each detector fibre.  The associated coherence length is then longer than the
longest expected pulse duration thus ensuring high visibility interference.

\begin{figure}
\epsfxsize=17cm
\epsffile{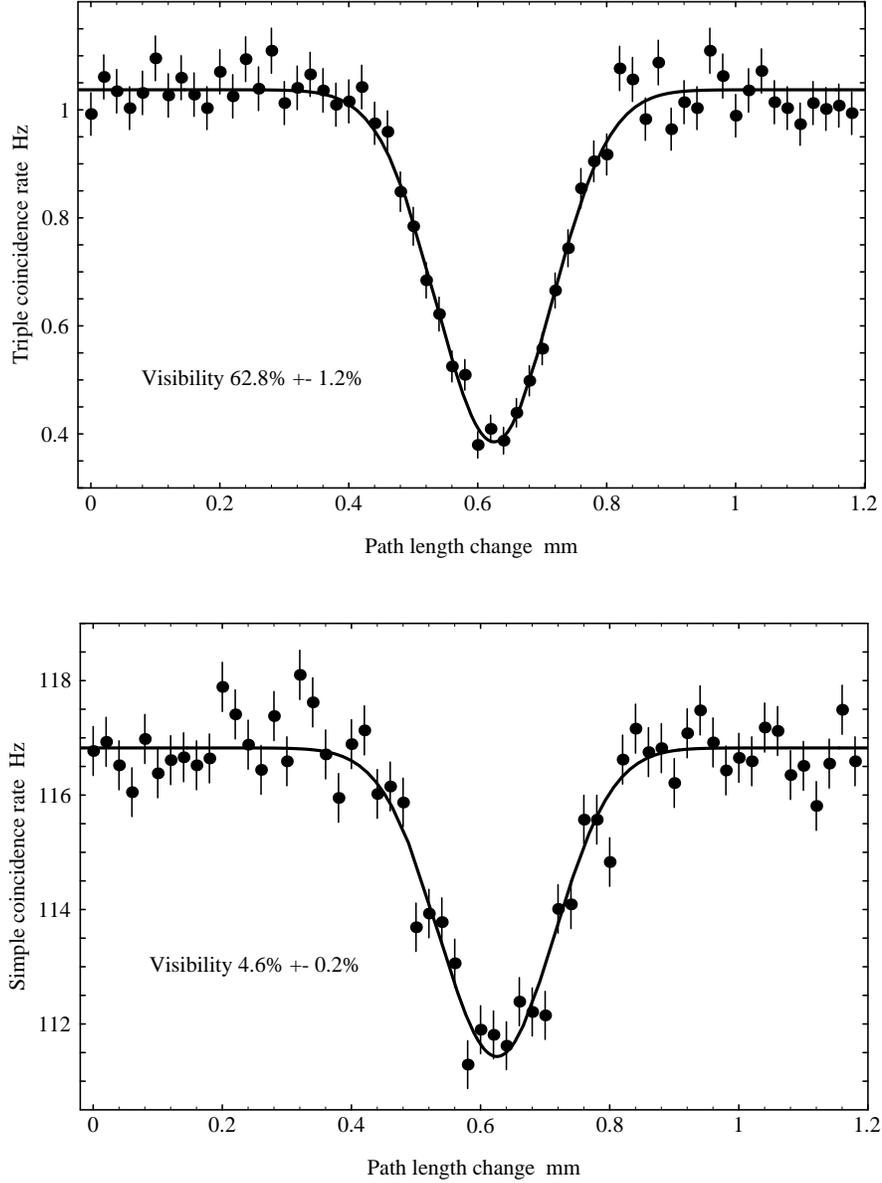}
\caption{ (a) Triple coincidence rate measured as a function of
delay $\Delta X$ in the weak laser path. Solid line shows a least
squares fit based on the Gaussian function in eq 8
(b) Ungated coincidence rate as a function of delay $\Delta X$.}
\label{result}
\end{figure}

In the experiment the counting rates due to downconverted photons were of
order 5 kilo-counts per second (kcps) while those from the laser were 103 kcps
(measured after the beamsplitter by sequentially blocking the laser and
downconversion source).  The two coincidence rates between between $c$- and
$d$-mode detectors and the gate detector each were $\simeq$500cps.  Doubling
these figures to obtain the rates before the beamsplitter allows an estimate
of the triple coincidence rate from equation 8 and the laser repetition rate
of 100MHz.  This predicts a triple coincidence rate of about 1.1 cps away from
overlap.  Measuring the coincidences only between $c$- and $d$-mode detectors
with the gating bypassed we expect a rate of 117cps if we assume coherent
statistics for the filtered mode-locked laser.  By simultaneously collecting
data with and without the gating active we can compare classical and
non-classical interference in the same experiment.  In a typical experiment
adequate statistics could only be obtained from at least 100 counts per point
leading to minimum experiment durations of several hours.  Experimental
results showing coincidence rates as a function of optical delay are seen in
figures 4.  Both gated and ungated results show a coincidence reduction at
pulse overlap which is well fitted by a Gaussian function of $1/e$ half width
133$\mu$m (443fs).  The visibility of the gated coincidence dip is
62.8$\pm$1.2\% while that for the ungated measurement is 4.6$\pm$0.2\%.

This visibility, being greater than 50\% is clear evidence of non-classical
interference.  It is reduced from 100\% partly by background triple
coincidences from small numbers of pair events arising in the laser and
downconversion pulses (about 0.1 events per second).  For 3nm full width half
maximum bandwidth filters equation 8 suggests we should measure 83$\mu$m
(277fs) $1/e$ half width.  This indicates that the pulse length of the
downconverted light has been stretched by the walk-off caused by group
velocity dispersion in the BBO crystals.  This further contributes to reduce
the visibility.  Initial experiments with a 3mm LiIO$_3$ crystal (with worse
walk-off) showed visibilities of 36\% and 200$\mu$m 1/e half width.  The
visibility of the classical interference is about half that expected from
equation 7 using measured count rates (9.2\%).  We expect that this
discrepancy is also due to walk-off effects.

To go beyond this experiment and generate multiparticle interference we will
need higher efficiency single photon (Fock) state generators and detectors.
Here we are limited to creating much less than one pair photon per mode per
pulse in the crystal to minimise the probability of detecting more than one
downconversion photon per pulse.  The maximum tolerable value will be around
$n_{dc}\leq 0.1$ pair-photon per pulse.  Similarly our effective detector
efficiencies are of order 10\% when using single mode fibre.  The coincidence
rate in an N-fold single photon interference/entanglement experiment where
each 1-photon state is selected by gating, will scale as $C\simeq
Rn_{dc}^N\eta^{2N}$ where R is the laser repetition rate.  If we gate only one
photon as here then this improves to $C\simeq Rn_{dc}^N\eta^{N+1}$.  This
exponential reduction of rate with increasing $N$ is the main factor limiting
the further development of quantum computers as it balances any exponential
increase in throughput.

In conclusion have shown here that when we take two seemingly
separate particles and mix them in such a way that they become
indistinguishable then we will see interference effects that are
manifestly non-classical.

We acknowledge helpful discussions with Prof F De Martini, Dr H Weinfurter and
D Wardle during the preparation of this work.

\end{document}